\documentclass[11pt]{article}
\usepackage{latexsym}
\usepackage{amsmath}
\usepackage{amssymb}
\usepackage{epsfig}

\usepackage{rotating}

\renewcommand{\baselinestretch}{2.0}
\newcommand{\be}{\begin{equation}}
\newcommand{\ee}{\end{equation}}
\newcommand{\ben}{\begin{eqnarray}}
\newcommand{\een}{\end{eqnarray}}

\newcommand{\eff}{ \mbox{\it{\Large e}} }

\newcommand\ind{\bot\hspace*{-6pt}\bot}
\newcommand\jind[2]{#1\ind#2}

\newcommand {\inispace}{\renewcommand{\baselinestretch}{2.0}}
\newcommand {\tabspace}{\renewcommand{\baselinestretch}{1.5}}
\newcommand {\vet}{\mbox{vec}}

\newcommand {\dn}[1] {\mbox {\boldmath$#1$} }

\newcommand {\etal} { {\it et al.} }

\newtheorem{definition}{{\sc Definition}}

\usepackage{color}

\usepackage{graphicx}

\title{Bayesian Analysis of Marginal Log-Linear Graphical Models for Three Way Contingency Tables}
\author{Ioannis Ntzoufras and Claudia Tarantola }
\author{ Ioannis Ntzoufras \\ Department of Statistics,  Athens University of
Economics and Business,  Greece.
\\
\and Claudia Tarantola \thanks{Address for correspondence: Claudia
Tarantola, Department of Economics and Quantitative Methods,
University of Pavia, Via S. Felice 7, 27100 Pavia, Italy.
\texttt{E-mail:~claudia.tarantola@unipv.it}}\\
Department of Economics and Quantitative Methods, University of
Pavia, Italy
 }

\begin{document}

\small \maketitle \normalsize

\inispace

\begin{abstract}

This paper deals with the Bayesian analysis of graphical
models of marginal  independence for three way contingency tables.
We use  a marginal log-linear parame\-tri\-zation, under which the model is defined through suitable
zero-constraints on the interaction parameters calculated within
marginal distributions.
 We undertake a comprehensive Bayesian
analysis of these models, involving suitable choices of prior
distributions, estimation, model determination, as well as the
allied computational issues.  The methodology is illustrated with
reference to two real data sets.

\end{abstract}


\section{Introduction}

Graphical models of marginal independence  were originally
introduced by Cox and Wermuth (1993) for the analysis of multivariate
Gaussian distributions. They compose a family of multivariate
distributions incorporating the marginal independences
represented by a bidirected graph.
The nodes in the  graph correspond to a set of  random variables and the bidirected edges
represent the pairwise associations between them. A missing edge
from a pair of nodes indicates that the corresponding variables
are marginally independent. The complete list of marginal
independences implied by a bidirected graph were studied by
Kauermann (1996) and by Richardson (2003) using the so-called
Markov properties.

The analysis of the Gaussian case can be easily performed both in  classical and
Bayesian frameworks
since marginal independences  correspond to zero
constraints in the variance-covariance matrix.
The situation is more complicated in the
 discrete case, where   marginal independences correspond to non linear constraints on the set of
parameters. Only recently  parameterizations for these models have
been proposed by Lupparelli (2006), Lupparelli \etal (2008) and
Drton and Richardson (2008). In this paper we use the
parameterization proposed by Lupparelli (2006) and Lupparelli
\etal (2008) based on the class of marginal log-linear models of
Bergsma and Rudas (2002). Each log-linear parameter is calculated
within the appropriate marginal distribution and a  graphical model of marginal independence
 is defined by zero constraints on specific higher
order log-linear parameters.
Alternative parameterizations have been proposed by Drton and Richardson (2008)
based  on the Moebius inversion and by Lupparelli \etal (2008) based on multivariate logistic
representation  of the models of  Glonek and McCullagh (1995).

 We present a comprehensive Bayesian analysis of
discrete  graphical models of mar\-gi\-nal independence, involving
suitable choices of prior distributions, estimation, model
determination as well as the allied computational issues. Here we
focus on the three way case where the joint probability of each
model under consideration can be appropriately factorized. We work directly in terms
of the vector of joint probabilities on which we impose the
constraints implied by the graph. Then we consider a minimal set
 of probability parameters expressing marginal/conditional independences and sufficiently
 describe the graphical model of interest.
  We introduce a conjugate prior distribution based on Dirichlet priors
 on the appropriate probability parameters. The prior distribution factorize
  similarly to the likelihood. In order to make the prior distributions `compatible'
   across models we define all probability parameters (marginal and conditional ones)
    of each model from the parameters of the joint distribution of the full table. In order to specify the prior parameters
    of the Dirichlet prior distribution, we adopt ideas based on the power prior approach of Ibrahim and Chen (2000) and Chen \etal (2000).

The plan of the paper is as follows. In Section 2 we introduce
graphical models of marginal independence, we establish the
notation, we present  Markov properties and we  explain in detail
their log-linear parameterization.
Section 3 illustrates a suitable factorization of the likelihood function  for all models of marginal independence in three-way tables.
In Section 4,
we consider conjugate prior distributions,
we present an imaginary data approach for prior specification
and we compare alternative prior set-ups.
Section 5 provides posterior model and parameter distributions which can be easily calculated via conjugate analysis.
Two illustrative examples are presented in Section 6.
Finally, we end up with a discussion and some final comments regarding our current research on the topic.

\section{Preliminaries}

\subsection{Graphical Models of Marginal Independence}

A   bidirected graph  $G=\left(\mathcal{V}, E\right)$ is
characterized by a vertex set ${\mathcal{ V}}$
and an edge set $E$ with the property that $(v_i,v_j) \in E$ if and only if  $(v_j,v_i) \in E$. 
We  denote each bidirected edge by
$(\overleftrightarrow{v_i,v_j}) = \big\{ (v_i, v_j), \, (v_j, v_i)
\big\}$ and we represent it with a
 bidirected arrow. If a vertex $v_i$ is adjacent
to another vertex $v_j$, then $v_j$ is said to be \textit{spouse}
of $v_i$, and we write $v_j \in sp(v_i)$. For a set $A \subseteq
{\cal V}$, we define $sp(A)= \cup \left(sp(v)|v \in A\right)$. The
degree $d(v)$ of a vertex $v$ is the cardinality of  the spouse
set. A path connecting two vertices, $v_0$ and $v_m$, is a finite
sequence of distinct vertices $v_0, \ldots v_m$ such that
$(v_{i-1},v_i), \; i=1,\ldots,m$, is an edge of the graph. A
vertex set $C \subseteq { \mathcal{V}}$ is connected if every two
vertexes $v_i$ and $v_j$ are joined by a path in which every
vertex is in C. Two sets $A,B \in \mathcal{V}$ are separated by a
third set $S \in \mathcal{V}$ if any path from a vertex in A to a
vertex in $B$ contains a vertex in $S$. It can be shown that,  if
a subset of the nodes $D$ is not connected then there exist a
unique partition of it into maximal (with respect to inclusion)
connected set $C_1, \ldots,C_r$ \be D=C_1 \cup C_2 \cup \hdots
\cup C_r. \label{DC} \ee

The graph is used to represent marginal independences between a
set of discrete random variables $X_{\cal V}= \big( X_v,~ v\in
{\cal V} \big)$, each one taking values $i_v \in {\cal I}_{v}$.
The cross-tabulation of variables $X_{\cal V}$ produces a
contingency table of dimension $|{\cal V}|$ with cell frequencies
${\dn n} = \Big( n(i), ~ i \in {\cal I} \Big)$ where ${\cal
I}=\times_{v\in {\cal V}}{\cal I}_v$. Similarly for any marginal
$M \subseteq {\cal V}$, we denote with $X_{M}=\big( X_v,~v\in
M\big)$ the set of variables which produce the marginal table with
frequencies ${\dn n}_M = \Big( n_M(i_M),~i_M \in {\cal I}_M\Big)$
where ${\cal I}_M=\times_{v\in M}{\cal I}_v$. The marginal cell
counts are the sum of specific elements of the full table and are
given by

$$
n_{M}(i_M) = \sum_{ j \in {\cal I}_{\overline{M}|i_{M}}} n(j)
$$

\noindent where $\overline{M}={\cal V} \setminus M$ and ${\cal
I}_{\overline{M}|i_{M}}= \{ j \in {\cal I}:
j_{M}=i_{M} \}$. Therefore ${\cal I}_{\overline{M}|i_M}$ refers to
all cells of the full table for which the variables of the $M$
marginal are constrained to the specific value $i_M$.

A graphical model of marginal independence is constructed via the following Markov properties.
%
%
\begin{definition}: \underline{Connected Set Markov Property} (Richardson, 2003).
The distribution of a random vector
$X_{\cal V}$ is said to satisfy the connected set Markov property if
 \be \jind{X_C}{X_{\mathcal{V}\setminus(C \cup sp(C))}} \label{CSMP} \ee
 whenever $\emptyset \neq C \subseteq {\cal
V}$ is a connected set.
\end{definition}

A more exhaustive Markov property is the global Markov property,
which requires all the marginal independences in (\ref{CSMP}), but
also additional conditional independences.

\begin{definition}: \underline{Global Markov Property} (Kauermann, 1996 and Richardson, 2003).
The distribution of
a random vector  $X_{\cal V}=\{X_v, v \in { V}\}$ satisfies the
Global Markov property if \be A ~\mbox{is separated from}~ B
~\mbox{by} ~ V\setminus(A \cup B \cup C) ~\mbox{in} ~ G
~\mbox{implies}~  \jind{X_A}{X_{B}|X_C} \label{GMP}, \ee with $A$,
$B$ and $C$ disjoint subsets of $V$, and C may be empty.
\end{definition}

Despite the global
Markov property is more exhaustive (in the sense that indicates
both marginal and conditional independences), Drton and Richardson
(2008) pointed out that a distribution satisfies the global Markov
property if and only if it satisfies the connected set Markov
property.

>From the global Markov property, we directly derive that if two
nodes $i$ and $j$ are disconnected, then $\jind{X_i}{X_j}$ that is
the variables are marginal independent. The same is true for any
two sets $A\subset {\cal V}$ and $B\subset {\cal V}$ that are
disconnected, implying that $\jind{A}{B}$ (are marginal
independent). This can be easily generalized  for any given
disconnected set $D$ satisfying (\ref{DC}). Then the  global
Markov property for the bidirected graph $G$ implies
$$\jind{X_{C_1}}{X_{C_2}}\jind{}{} \hdots \jind{}{X_{C_{r}}}.$$

According to Drton and Richardson (2008), a  discrete marginal
 graphical model, associated to a bidirected graph G,
is a family $P(G)$  of joint distributions for a categorical
random vector $X_{\cal V}$ satisfying the global Markov property
(or equivalently the connected set Markov property). Following the
above, for every not connected set $D\subseteq {\cal V}$, it holds
that
\begin{equation}
P(X_D=i_D)= \prod_{k=1}^{r} P(X_{C_k} = i_{C_k})
\label{factorization}
\end{equation}
where $C_1, \ldots,C_r$ are the inclusion maximal connected sets satisfying (\ref{DC}).


\subsection{A Parameterization for Marginal Log-Linear Models}

Lupparelli (2006) and Lupparelli \etal (2008) show that it is possible
to define a parameterization for any set $X_{\cal V}$ of
categorical variables, by using the marginal log-linear model by
Bersgma and Rudas (2002).

Bergsma and Rudas (2002) suggested to  work in terms of log-linear
parameters ${\dn \lambda}$ obtained from a specific set of
marginal tables. They consider the following model \be
\dn{\lambda} = \dn{C} \log \Big( \dn{M} \vet(\dn{\pi}) \Big)
\label{lambda_of_marginal_model} \ee where ${\dn \pi} = \Big(
\pi(i), ~i\in{\cal I}\Big)$ is the joint probability distribution
of $X_{\cal V}$ and $\mbox{vec}(\dn{\pi})$ is a vector of
dimension $|{\cal I}|$ obtained by rearranging the elements
$\dn{\pi}$ in a reverse lexicographical ordering of the
corresponding variable levels with the level of the first variable
changing first (or faster). For example in a $2\times 2$ table the
vector of probabilities will be given by
 $\mbox{vec}(\dn{\pi})= \Big( \pi( 1,1), \pi(2,1),\pi(1,2),\pi(2,2)\Big)^T$. In this paper
we assume that the parameter vector $\dn{\lambda}$ satisfies
sum-to-zero constraints and  we indicate with $\dn{C}$  the
corresponding contrast matrix. Finally $\dn{M}$ is the
marginalization matrix which specifies from which marginal we
calculate each element of $\dn{\lambda}$. An algorithm for
constructing $\dn{C}$ and $\dn{M}$ matrices is given in the
Appendix ( for additional details see Appendix A in Lupparelli,
2006).

\subsubsection{Properties of Marginal Log-Linear Parameters}

Let $M \subseteq {\cal V}$ be a generic marginal, and indicate with $S({ M})$  the class of all subsets of ${ M}$ and with  $\dn{\cal E}_{M} \in S({ M})$
 the set of effects obtained from marginal $M$.

Given ${\cal M}=\{ M_1, M_2, \dots, M_{|{\cal M}|} \}$
the set of marginals used to calculate the log-linear parameters ${\dn \lambda}$,
 we  denote by
$\dn{\lambda}_{\eff}^M = \Big( \dn{\lambda}_{\eff}^M(i_{\eff}), ~i_{\eff} \in {\cal I}_{\eff},~M\in{\cal M} \Big)$,
the set of parameters for effect $\eff \subseteq M$ estimated by the marginal $M$
and by $\dn{\lambda}^M$ the set of all parameters estimated by the same marginal.

According to Bersgma and Rudas (2002), in order to obtain a well-defined parameterization,
it is important to allocate the interaction parameters $\dn{\lambda}$ among
 the chosen marginals to get a {\it complete} and {\it hierarchical} set of parameters.

\begin{definition}: \underline{Complete and hierarchical set of parameters}.
A  set of marginal log-linear parameters ${\dn \lambda}$ is called hierarchical and complete if:
\begin{itemize}
\item[i)]           The  elements of ${\cal M}$ are ordered in a non decreasing order,
          which means that no marginal is a subset of any preceding one,
          i.e. $ {M}_i \not \subseteq { M}_j \;\; \mbox{if~~} i>j $
          (hierarchical ordering).

          Furthermore, we require the last marginal in the sequence to be the set all vertices under
          consideration, therefore $M_{|{\cal M}|}={\cal V}$ .

\item[ii)] From each marginal ${M}_i$ we can calculate only the
            effects that was not possible to get from the ones preceding it in the ordering under consideration;
            hence the sets ${\cal E}_{M_i}$ of the effects under consideration are given by
            $$ {\cal E}_{M_1}=S({M}_1)  \;\; \mbox{and} \;\;
               {\cal E}_{M_i}=S({ M}_i)  \setminus \left\{ \cup_{k=1}^{i-1} S({ M}_k) \right\} \;
               \mbox{for} \;\; i=1,\ldots,|{\mathbb M}|~.
            $$
\end{itemize}
\end{definition}

The above set of parameters define a parametrization of the distribution on the contingency
table (see Bergsma and Rudas, 2002, for a formal definition).
It is called complete since each parameter is estimated  from one and only one marginal and
hierarchical because the full set of parameters is generated by marginals in a non decreasing ordering
(Bersgma and Rudas, 2002, p 143-144).

For every complete and hierarchical set of parameters,
the inverse transformation of (\ref{lambda_of_marginal_model}) always
exists but it cannot be analytically calculated (Lupparelli, 2006, p. 39).
Iterative procedures have been used to calculate $\dn{\pi}$
for any given values of $\dn{\lambda}$ and hence the likelihood of the
graph under consideration (Rudas and Bergsma, 2004 and Lupparelli, 2006).

An important problem when working with marginal log-linear parameters ${\dn \lambda}$
is that we may end up with the definition of non-existing joint marginal probabilities.
A way to avoid this is to consider the parameters' variation independence property;
see Bergsma and Rudas (2002).
A set of parameters is variation independent when the range of possible values of one of them does not depend
on the other's value.
Hence  the joint range of the parameters is the Cartesian product of the separate ranges of the parameters involved.
This property ensures the existence of a common joint distribution deriving from the marginals of the model under consideration.
Moreover, variation independence ensures strong compatibility which implies both
compatibility of the marginals  and  existence of a common joint distribution.
Compatibility of the marginals
 means that from different distributions we will end up to the same distribution for the common
parameters, for example from marginals AB and AC we get the same marginal for A.
To assure variation independence of hierarchical log-linear parameters,
a generalization of the classical decomposability concept is needed.

\begin{definition}:
 \underline{Decomposable set of Marginals}.  A class of incomparable
 (with respect to inclusion) marginals ${\cal M}$ is called
decomposable if it has \underline {at most two} elements
($|{\cal M}|\leq 2$) or if there is an ordering ${ M}_1,
\ldots, { M}_{|{\cal M}|}$ of its elements such that, for
$k=3,\ldots, |{\cal M}|$ there exist at least one $j_k<k$ for
which the running intersection property is satisfied, that is
$$
\left(\cup_{i=1}^{k-1}{ M}_i \right)\cap { M}_k={ M}_{j_k}\cap { M}_k  ~.
$$
\end{definition}

\begin{definition}: \underline{ Ordered decomposable set of marginals}. A class of marginals ${\cal M}$ is ordered decomposable if
it has at most two elements or if there is a hierarchical ordering
${ M}_1, \ldots, {M}_{|{\cal M}|}$ of the marginals and, for
$k=3,\ldots, |{\cal M}|$, the maximal elements (in terms of inclusion)
of $\{{ M}_1,\ldots,{ M}_k\}$ form a decomposable set.
\end{definition}



The importance of the above property is due to the theorem 4 of Bergsma and Rudas (2002)
where they proved that a set of complete and hierarchical marginal log-linear parameters is variation independent if and only if
the ordering of the marginals involved is ordered decomposable. 
Hence order decomposability ensures the existence of a well defined joint probability.

\subsection{Construction of  Marginal Log-Linear Graphical Models}

Based on the results of the previous subsection,  Lupparelli
(2006) and  Lupparelli \emph{et al.} (2008)  proposed a strategy
to construct a marginal log-linear parametrization for the family
of discrete bidirected graphical.

Initially we need to consider a
set of a parameters ${\dn \lambda}$ derived from a hierarchical ordering
 of the  marginals in ${\cal D}(G) \cup {\cal
V}$; where ${\cal D}(G)$ is the set of all disconnected components
of  the bidirected graph  ${G}$. Then, we must set the highest
order log-linear interaction parameters of ${\cal D}(G)$ equal to
zero.

More precisely, we need to consider the  following steps:

\begin{enumerate}

\item [i)]  Construct a hierarchical ordering of the marginals
$M_i \in {\cal D}(G)$.

\item [ii)] Append the marginal $M = {\cal V}$ (corresponding to the full table under consideration)
            at the end of the list if it is not already included.

\item [iii)]  For every marginal table $M_i\in {\cal D}(G)\cup{\cal V}$
            estimate all parameters of effects in $M_i$ that have not already estimated from the preceding marginals.

\item [iv)] For every marginal table $M_i \in {\cal D}(G)$, set the highest order log-linear interaction parameter equal to zero; see Proposition 4.3.1 in
Lupparelli (2006).

\end{enumerate}

In the three way case the log-linear parameters the marginals are always obtained from a set of order of order decomposable
 marginals, hence the parameters are variation independent.


\section{Likelihood Decomposition}
\label{likelihood}

 In this paper we propose to use a different approach from the one
by   Rudas and Bergsma (2004) and Lupparelli (2006) in order to
estimate the joint distribution ${\dn \pi}$ of a graph $G$. We
 impose the constraints  implied
by the graph $G$ directly on the joint probabilities ${\dn \pi}$.
We work with a minimal set of probability parameters $\dn{\pi}^G$
expressing marginal/conditional independences and sufficiently
describe the graphical model $G$ under investigation. {By this
way} we can always reconstruct the joint distribution ${\dn \pi}$
for a given graph $G$ via $\dn{\pi}^G$ and  then simply calculate
the marginal log-linear parameters directly using
(\ref{lambda_of_marginal_model}).  Here we focus on the three
way case where the joint probability of each model can be
appropriately factorized for any graph $G$.

\begin{figure}[h]
\centering
\begin{tabular}{cccc}
\includegraphics[width=3cm]{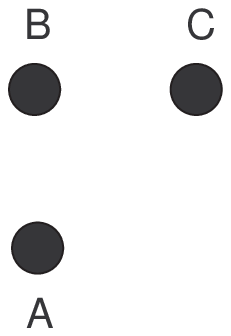} &
\includegraphics[width=3cm]{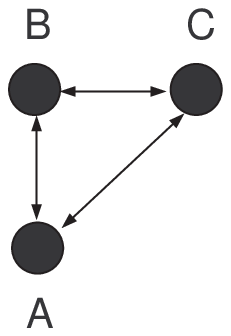} &
\includegraphics[width=3cm]{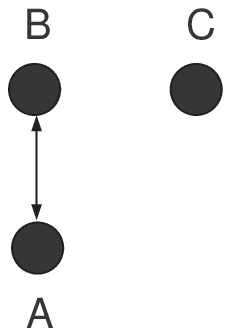} &
\includegraphics[width=3cm]{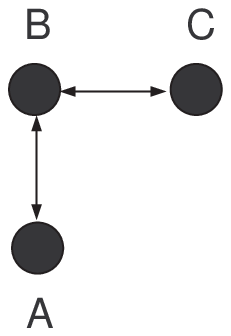} \\
(a) Independence Model       & (b) Saturated Model & (c) Edge Model
& (d) Gamma Model
\end{tabular}
\caption{Type of Graphs in Three Way Tables}
\label{graphs}
\end{figure}

For every three way contingency table eight possible graphical models models exist
which can be represented by four different types of graphs:
the independence, the saturated, the
edge and the gamma structure graph (see Figure \ref{graphs}). The
independence graph is the one with the empty edge set
($E=\emptyset$), the saturated is the one containing all possible
edges $\left[ E=\big( \overleftrightarrow{(v,w)}: v\neq w \in
{\cal V} \big) \right]$, an edge graph is the one having only one
single edge  and a gamma structure graph is one represented by a
single path of length two. In a three way table, three `edge' and
three `gamma' graphs are available.

For the saturated model $G_S$, we get all parameters from the full table, i.e. $\dn{\pi}^{G_S} = \dn{\pi}$.
Thus, the likelihood is directly written as
\begin{eqnarray*}
f( \dn{n} | \dn{\pi}, G_S ) = \frac{\Gamma(N+1)} {\prod \limits _{i \in {\cal I}} \  \Gamma\Big(n(i)+1\Big)} \prod_{i \in {\cal I}} \pi(i)^{n(i)}
\end{eqnarray*}
where $N=\sum_{i \in {\cal I}} n(i)$ is the total sample size.

The joint distributions for the independence and the edge models can be easily expressed using
the equation
\begin{eqnarray*}
f(\dn{n}|\dn{\pi}^G, G) =\frac{\Gamma(N+1)}{\prod\limits _{i \in {\cal I}}\ \Gamma\Big(n(i)+1\Big)}
\prod_{ d \in {\cal D}(G)} \; \prod_{i_d \in {\cal I}_d}\pi_d(i_d)^{n(i_d)}~
\end{eqnarray*}
where
$\dn{\pi}^G = \Big( \dn{\pi}_d, \, d\in {\cal D}(G)\Big)$
and
$\dn{\pi}_d = \Big( \pi_d(i_d), \, i_d\in {\cal I}_d \Big)$ is the vector of  probabilities corresponding to the table of the marginal $d$.
The above equation derives directly  from (\ref{factorization}) since the graph is disconnected.

Finally, the decomposition of the gamma structures is not as straightforward as in the previous
case. In addition to specific marginal probability parameters we also need to use
some conditional ones.
Let us denote by ``$c$'' the corner node, that is the vertex with degree 2 and
by $\overline{c}={\cal V} \setminus c$ the end-points of the path.
The set of disconnected marginals is equal to the end-point vertices of the graph,
hence $\overline{c}={\cal D}(G)$.
Then the likelihood can be written as
{\small
\begin{eqnarray*}
f(\dn{n}|\dn{\pi}^G,G) = \frac{\Gamma(N+1)}{\prod \limits _{i \in {\cal I}}\ \Gamma\Big(n(i)+1\Big)}
\prod_{i_{\overline{c} }\; \in
{\cal I}_{\overline{c} }} \left(\prod_{i_{c  \in {\cal I}_c} }
\pi_{c|\overline{c}}(i_{c}|i_{\overline{c}})^{n(i_{c},i_{\overline{c}})}\right)
\left[ \prod_{ d \in {\cal D}(G)} \; \prod_{i_d \in {\cal I}_d}  \pi_d(i_d)^{n(i_d)}\right]~,
\end{eqnarray*}
} where $\dn{\pi}^G = \Big( \dn{\pi}_{c|\overline{c}}, \,
\dn{\pi}_d, \, d\in {\cal D}(G) \Big)$, here ${\cal D}(G)={\cal V}
\setminus c$  and
$$
\dn{\pi}_{c|\overline{c}} = \Big(
\pi_{c|\overline{c}}(i_{c}|i_{\overline{c}}),~i_{c} \in {\cal
I}_{c}, ~i_{\overline{c}} \in {\cal I}_{\overline{c}} \Big)
$$
are all the conditional probabilities of $c$ given $\overline{c}$.

The above factorization can be easily adopted to get maximum likelihood
estimates analytically and avoid the iterative procedure used by  Rudas and Bersgma (2004)  and Lupparelli (2006).
In this paper we work using conjugate priors
on the appropriate probability parameters of the above parametrization
and then calculate the corresponding log-linear parameters.

\section{Prior distributions on cell probabilities}
\label{priors}

%

\subsection{Conjugate Priors}

\label{conjugate_prior}

For the specification of the prior distribution of the probability
parameter vector we initially consider a Dirichlet distribution
with parameters ${\dn \alpha}=\big( \alpha(i), \, i \in {\cal I} ~
\big)$ for the vector of the joint probabilities of the full table
${\dn \pi}=\big( \pi(i), \, i \in {\cal I} \big)$, where ${\cal
I}$ is the set of all cells of the table under consideration.
Hence, for the full table the prior density is given by \be f(
{\dn \pi} ) = \frac{\varGamma(\alpha) }{ \prod \limits_{i\in
  {\cal I}}\  \varGamma\Big(\alpha(i)\Big) } \prod_{i \in{\cal I}} \pi(i)^{\alpha(i) - 1}
= f_{{\cal D}i}( \dn{\pi}; \, \dn{\alpha} )
\label{prior_full_table}
\ee
where $f_{{\cal D}i}\big( \dn{\pi} ;~\dn{\alpha} \big)$
is the density function of the Dirichlet distribution evaluated at $\dn{\pi}$ with parameters $\dn{\alpha}$
and $\alpha=\sum_{i\in{\cal I}} \alpha(i)$.

Under this set-up, the marginal prior of $\pi(i)$ is a Beta distribution
with parameters $\alpha(i)$ and $\alpha-\alpha(i)$,
i.e. $\pi(i) \sim Beta\Big(\alpha(i),\alpha-\alpha(i)\Big)$.
The prior mean and variance of each cell is given by
$$
E\big[ \pi(i) \big] = \frac{\alpha(i)}{\alpha} \mbox{~~and~~}
V\big[ \pi(i) \big] = \frac{\alpha(i)
\{\alpha-\alpha(i)\}}{\alpha^2(\alpha+1)}~.
$$
When no prior information is available then we usually set all
$\alpha(i)=\frac{\alpha}{|{\cal I}|}$ resulting to
$$
E\big[ \pi(i) \big] = \frac{1}{|{\cal I}|} \mbox{~~and~~} V\big[
\pi(i) \big] = \frac{  |{\cal I}| - 1 }{ |{\cal I}|^2 (\alpha+1)
}~.
$$

Small values of $\alpha$ increase the variance of each cell probability parameter.
Usual choices for $\alpha$ are the values $|{\cal I}|/2$ (Jeffrey's prior),  $|{\cal I}|$ and  $1$
(corresponding to $\alpha(i)$ equal to $1/2$, $1$ and $1/|{\cal I}|$ respectively);
for details see Dellaportas and Forster (1999).
The choice of this prior parameter value is of prominent importance
for the model comparison due to the well known sensitivity of the
posterior model odds and the Bartlett-Lindley paradox (Lindley, 1957, Bartlett, 1957).
Here this effect is not so adverse, as for example in usual variable selection for generalized linear models,
for two reasons. Firstly even if we consider the limiting case
where $\alpha(i)=\frac{\alpha}{|{\cal I}|}$ with $\alpha\rightarrow 0$,
 the variance is finite and equal to $(|{\cal I}| - 1)/|{\cal I}|^2$.
Secondly, the distributions of all models are constructed from a common
distribution of the full model/table making the prior distributions `compatible'
across different models (Dawid and Lauritzen, 2000 and Roverato and Consonni, 2004).

The model specific prior distributions are defined by the
constraints imposed by the model's graphical structure and the
adopted  factorization. The prior distribution also factorizes in
same manner as the likelihood described in section
\ref{likelihood}. Thus, the prior for the saturated model is the
usual Dirichlet (\ref{prior_full_table}).

For the independence and edge models the prior is given by
\begin{eqnarray*}
f \left( \dn{\pi}^G \big| G \right) &=& \prod_{ d \in {\cal D}(G)}
\; \left[ \frac{\Gamma\left( \alpha \right)}
                    {\prod \limits_{i_d \in {\cal I}_d} \
                    \Gamma\Big(\alpha(i_d)\Big)}  \prod_{i_d \in {\cal I}_d} \pi_d(i_d)^{\alpha{(i_d)-1}}\right] 
                =\prod_{ d \in {\cal D}(G)} \;
                    f_{{\cal D}i}( \dn{\pi}_d;~\dn{\alpha}_d ) \\
\end{eqnarray*}
with parameter vector $\dn{\pi}^G = \big( \dn{\pi}_d, \, d\in {\cal D}(G)\big)$.
We denote the above density which is a simple product (over all
disconnected  sets) of Dirichlet distributions by
\begin{eqnarray}
f\left( \dn{\pi}^G \right)= f_{\cal PD} \Big( \dn{\pi}_d;
~\dn{\alpha}_d, \, d\in {\cal D}(G) \Big). \label{prior_edge}
\end{eqnarray}

For the gamma structure the prior is given by
\begin{eqnarray}
p\left( \dn{\pi}^G \right) &=& \prod_{i_{\overline{c} }\; \in
{\cal I}_{\overline{c} }} \left\{
\frac{\Gamma\Big(\alpha(i_{\overline{c}}) \Big)} {\prod
\limits_{i_{c} \in {\cal I}_{c}}\
\Gamma\Big(\alpha(i_{c},i_{\overline{c}})\Big)} \left(\prod_{i_{c
\in {\cal I}_{c}} }
\pi_{c|\overline{c}}(i_{c},i_{\overline{c}})^{\alpha(i_{c},i_{\overline{c}})-1}\right)  \right\} \label{prior_gamma} \\
&\times&   ~f_{\cal PD} \Big( \dn{\pi}_d; ~\dn{\alpha}_d, \, d\in
{\cal D}(G) \Big) \nonumber
\end{eqnarray}
with parameter vector
$\dn{\pi}^G \big( \dn{\pi}_{c|\overline{c}}, \, \dn{\pi}_d, \, d\in {\cal D}(G)\big)$.
The fist part of equation (\ref{prior_gamma}), that
is the product for all level of $\overline{c}$ of Dirichlet
distributions of the conditional probabilities, can be denoted by
$f_{\cal CPD} \big( \dn{\pi}_{c|\overline{c}} \, ; ~\dn{\alpha}
\big)$. Then, the prior density  (\ref{prior_gamma}) can be
written as
\begin{eqnarray}
f\left( \dn{\pi}^G \right) &=& f_{\cal CPD} \big(
\dn{\pi}_{c|\overline{c}} \, ; ~\dn{\alpha}     \big) f_{\cal PD}
\big( \dn{\pi}_d; ~\dn{\alpha}_d, \, d\in {\cal D}(G) \big) ~.
\end{eqnarray}

In order to make the prior distributions `compatible' across
models, we define the prior parameters of $\dn{\pi}^G$
 from the corresponding parameters of the prior distribution
(\ref{prior_full_table}) imposed on the probabilities $\dn{\pi}$
of the full table; see Dawid and Lauritzen (2000), Roverato and
Consonni (2004).

Let us consider a marginal $M
\in {\cal M}(G)$ for which we wish to estimate the probability
parameters $\dn{\pi}_M = ( \pi_M(i_M), i_M \in {\cal I}_M  )$. The
resulting prior is $\dn{\pi}_M \sim {\cal D}i\big( \dn{\alpha}_M
\big)$, that is  a Dirichlet distribution with parameters ${\dn
\alpha}_{M} = \left( \alpha_{M}(i_M), i_M \in {\cal I}_{M}
\right)$ given by
$$
\alpha_{M}(i_M) = \sum_{ j \in {\cal I}_{\overline{M} \, |i_M}} \alpha(j),
$$
 see (i) of Lemma 7.2 in Dawid and Lauritzen (1993, p.1304).

For example, consider a three way table with ${\cal V}=\{ A, B, C\}$ and the marginal $M=C$.
Then the prior imposed on the parameters $\dn{\pi}_C$ of the marginal $C$
is given by
$$
\dn{\pi}_C \sim {\cal D}i\big( \dn{\alpha}_C \big) \mbox{~~with~~}
\alpha_C(i_C) = \sum_{i_A=1}^{|{\cal I}_A|}\sum_{i_B=1}^{|{\cal I}_B|} \alpha( i_A, i_B, i_C )
\mbox{~for~} i_C=1,2,\dots,|{\cal I}_C|,
$$
where
$\dn{\pi}_C =  \Big( \pi_C(1), \dots, \pi_C(|{\cal I}_C|)  \Big)$.

For the conditional distribution of $M_1|M_2$ with $M_1 \ne M_2 \in {\cal M}(G)$ we work in a similar way.
The vector
$\dn{\pi}_{M_1|M_2} ({\cdot}|i_{M_2}) = \big( \pi_{M_1|M_2}(i_{M_1}|i_{M_2}), i_{M_1} \in {\cal I}_{M_1} \big)$
a priori follows a  Dirichlet distribution
$$
\dn{\pi}_{M_1|M_2} ({\cdot}|i_{M_2})
\sim {\cal D}i
\Big( \dn{\alpha}_{ M_1 \cup M_2}(i_{M_1\cup M_2}), \, i_{M_1\cup M_2} \in {\cal I}_{M_1|i_{M_2}} \Big).
$$
The above structure derives from the decomposition of a Dirichlet as a ratio of Gamma distributions;
 see also Lemma 7.2 (ii) in Dawid and Lauritzen (1993, p.1304).

For example, consider marginals $M_1=A$ and $M_2=B$ in a three way contingency table with ${\cal V}=\{A,B,C\}$.
Then, for a specific level of variable $B$, say $i_B=2$,
$$
\dn{\pi}_{A|B}( \cdot | i_B=2 ) \sim {\cal D}i \big( \dn{\alpha}_{AB}(\cdot,2) \big)
$$
where
$\dn{\alpha}_{AB}(\cdot,2) = \Big( \alpha_{AB}(1,2), \alpha_{AB}(2,2), \dots, \alpha_{AB}( |{\cal I}_A|,2)\Big)$
and
$$
\alpha_{AB}(i_A,2) = \sum_{i_C=1}^{|{\cal I}_C|} \alpha_{ABC}(i_A,2,i_C).
$$

\subsection{Specification of Prior Parameters Using Imaginary Data.}

In order to specify the prior parameters of the Dirichlet prior
distribution, we adopt ideas based on the power prior approach of Ibrahim and Chen (2000) and Chen \etal (2000).
We use their approach to advocate sensible values for the Dirichlet prior parameters
on the full table and the corresponding induced values for the rest of the graphs as
described in the previous sub-section.
Let us consider imaginary set of data represented by the frequency table $\dn{n}^*=( n^*(i), i\in{\cal I} )$
of total sample size $N^* = \sum_{i \in {\cal I}} n^*(i)$
and a Dirichlet `pre-prior' with all parameters equal to $\alpha_0$.
Then the unnormalized  prior distribution can be obtained by the product of
the likelihood of $\dn{n}^*$ raised to a power $w$ multiplied by the `pre-prior' distribution.  Hence
\begin{eqnarray}
f( \dn{\pi} ) & \propto & f( \dn{n}^* |\dn{\pi} )^w \times ~
                f_{{\cal D}i}\big( \dn{\pi} ; ~\alpha(i)=\alpha_0, i \in {\cal I}
                \big)\nonumber\\
               &   \propto &  \prod_{i \in {\cal I}} \pi(i)^{w \, n^*(i) + \alpha_0 -1} \nonumber\\
              & =& f_{{\cal D}i}\big( \dn{\pi} ;~\alpha(i)=w \, n^*(i) + \alpha_0, i \in {\cal I}   \big)~.
              \label{power_prior}
\end{eqnarray}

Using the above prior set up, we expect a priori to observe a total
number of $w \, N^* + |{\cal I}|\alpha_0$ observations. The
parameter $w$ is used to specify the steepness of the prior
distribution and the weight of belief on each prior observation.
For $w=1$ then each imaginary observation has the same weight as
the actual observations.
Values of $w<1$ will give less weight to each imaginary observation while
$w>1$ will increase the weight of believe on the prior/imaginary data.
Overall the prior  will account for the
$ (w\,N^*+|{\cal I}|\alpha_0)/(w\,N^*+N+ |{\cal I}|\alpha_0) $
of the total information used in the posterior distribution. Hence for $w=1$,
$N^*=N$ and $\alpha_0 \rightarrow 0$ then both the prior and data
will account for $50\%$ of the information used in the posterior.

For $w=1/N^*$ then $\alpha(i)=p^*(i)+\alpha_0$ with $p^*(i)=n^*(i)/N^*$, the prior
 data $\dn{n}^*$ will account for information of one data point
while the total weight of the prior will be equal to
$(1+|{\cal I}|\alpha_0)/(1+N+|{\cal I}|\alpha_0)$.
If we further set $\alpha_0=0$, then the prior distribution (\ref{power_prior})
will account for information equivalent to a single observation.
This prior set-up will be referred in this paper as the unit information prior (UIP).
When no information is available, then we may further consider the choice
of equal cell frequencies $n^*(i)=n^*$ for the imaginary data in
order to support the simplest possible model under consideration.
Under this approach $N^* = n^* \times |{\cal I}| $ and
$w=1/N^* = \frac{1}{ n^*\times  |{\cal I}| } $ resulting to
$$
\dn{\pi} \sim {\cal D}i \Big( \alpha(i)= 1/|{\cal I}|, \, i \in {\cal I}  \Big)~.
$$
The latter prior is equivalent to the one advocated by Perks
(1947). It has the nice property that the prior on the marginal
parameters does not depend on the size of the table; for example,
for a binary variable, this prior will assign a $Beta(1/2,~1/2)$
prior on the corresponding marginal regardless the size
of the table we work with (for example if we work with $2^3$ or $2\times 4 \times 5 \times 4$ table).
This property is retained for any
prior distribution of type (\ref{power_prior}) with $w^*=1/N^*$,
$p^*(i)=1/|{\cal I}|$ and $\alpha_0 \propto 1/|{\cal I}|$.

\begin{sidewaystable}[hp!]
\caption{Table of Prior Variance in Comparison to the Unit
Information Prior (last row of the table)} \label{prior_params}
\begin{center}
\scriptsize
\begin{tabular}{l@{~~~}l@{~~~}l@{~~~~}l@{~~~~}l@{~~~}l}
\hline
&                           &            & \multicolumn{3}{c}{ Variance Ratio } \\
                                                                                \cline{4-6}
&Parameter                  & $V[\pi(i)]$ &  General Equation & $2\times 2 \times 2$ & $3\times 2 \times 4$ \\
\hline Jeffrey's &$\alpha(i)=1/2$            & $ 2~\frac{  |{\cal I}| - 1 }{ |{\cal I}|^2 \{|{\cal I}|+2\} }$ & $ \frac{4}{\Big(
|{\cal I}|+2 \Big)}$    & $0.4$ & $0.15$  \\ \vspace{0.2cm} Unit Exp. Cell&$\alpha(i)=1$              & $ \frac{  |{\cal I}| - 1
}{ |{\cal I}|^2 \{|{\cal I}|+1\}}$ &  $\frac{2}{ \Big( |{\cal I}|+1 \Big)}$& $0.22$ & $0.08$  \\\vspace{0.2cm}
 EBP       &
$\alpha(i)=p(i)$ & $\frac{1}{2}p(i) \Big( 1-p(i)\Big)$
          & $ \frac{  |{\cal I}|^2 }{  |{\cal I}| - 1 }  V[\pi(i)] $
          & $ 9.14 \times \Big( V[\pi(i)] \Big)$
          & $25.04 \times \Big( V[\pi(i)] \Big)$ \\ \vspace{0.2cm}
UIP       & $\alpha(i)=p^*(i)$ & $\frac{1}{2}p^*(i) \Big( 1-p^*(i)\Big)$
          & $ \frac{  |{\cal I}|^2 }{  |{\cal I}| - 1 } p^*(i) \Big( 1-p^*(i)\Big) $
          & $ 9.14 \times p^*(i) \Big( 1-p^*(i)\Big)$
          & $25.04 \times p^*(i) \Big( 1-p^*(i)\Big)$ \\

PP with $w=1/N^*$       & $\alpha(i)=p^*(i)+\alpha_0$ &
$S(p^*(i),\alpha_0,|I|)$\footnote{$S(p^*(i),\alpha_0,|I|)=\frac{\big(p^*(i)+\alpha_0
\big)\big(1+|I|\alpha_0-p^*(i)-\alpha_0\big)}{\big(1+|I|\alpha_0\big)^2\big(2+|I|\alpha_0\big)}$
}
          & $ \frac{ 2 |{\cal I}|^2 }{  |{\cal I}| - 1 } S(p^*(i),\alpha_0,|I|)
$
          & $ 18.28 \times S(p^*(i),\alpha_0,8)$
          & $50.08 \times S(p^*(i),\alpha_0,24)$ \\

\vspace{0.2cm}
UIP-Perks' &$\alpha(i)=1/|{\cal I}|$   & $ \frac{  |{\cal I}| - 1 }{  2|{\cal I}|^2} $ & $1$  & $1$ & $1$\\
\hline
\end{tabular}
\end{center}
\normalsize
\end{sidewaystable}

\subsection{Comparison of Prior Set-ups}

Since Perks' prior (with $\alpha(i)=1/|{\cal I}|$) has a unit information interpretation, it can be used as a
yardstick in order to identify and interpret the effect of any other prior distribution used.
Prior distribution with
$\alpha(i) < 1/|{\cal I}|$, or equivalently $\alpha<1$,
results in larger variance than the one imposed by our proposed unit information prior and hence
it a posteriori supports more parsimonious models.
On the contrary, prior distributions with  $\alpha(i) > 1/|{\cal I}|$,
or $\alpha>1$, result in lower prior variance and hence it a posteriori support models with more complicated graph structure.
So the variance ratio between  a Dirichlet prior with $\alpha(i)=\alpha/|{\cal I}|$
and Perks prior
is equal to
$$
VR=\frac{ V\big( \pi_i \big|~ \alpha(i)=\frac{\alpha}{|{\cal I}|}
\Big) }
     { V\big(\pi_i \big|~ \alpha(i)=|{\cal I}|^{-1} \big) }
= \frac{2}{\alpha+1}~.
$$

A comparison of the information used from some standard choices is
provided in Table \ref{prior_params}. From this Table, we observe
that Jeffreys' prior variance is lower than the corresponding
Perks' prior reaching a reduction of about $60\%$ and $85\%$ for a
$2^3$ and a $2\times 3\times 4$ table respectively. The reduction
is even greater for the prior of the Unit Expected Cell mean
 ($\alpha(i)=1$) reaching $78\%$ and $92\%$ respectively.

Finally, we use for comparison
an Empirical Bayes prior based on the UIP approach.
Hence we set the imaginary data $n^*(i)=n(i)$, $w=1/N$ and $\alpha_0=0$.
Then the resulting prior parameters are given by
$\alpha(i)= p(i)$, where $p(i)=n(i)/N$ is the sample proportion.
Under this set-up,
the prior variance for each $\pi(i)$ is equal to
$V[\pi(i)] = \frac{1}{2} p(i) \big( 1-p(i) \big)$.
Thus the above prior assumes that we have imaginary data with the same frequency table
as the observed one but they accounts for information equal to one data point (Empirical UIP).

\section{Posterior Model and Parameter Distributions  \label{post_distributions}}

Since the prior is conjugate to the likelihood the posterior can be derived easily as follows.
For the saturated model the posterior distribution is also a Dirichlet distribution
$\dn{\pi}\big|\dn{n}, G_S \sim {\cal D}i\big( \widetilde{\dn{\alpha}}  \big)$
with parameters
$$
\widetilde{\dn{\alpha}} =\big( \widetilde{\alpha}(i) = \alpha(i)+n(i), ~i \in {\cal I}\big).
$$
For the independence and the edge structure the density of the posterior distribution is
is equivalent to (\ref{prior_edge}),
$f(\dn{\pi}^G \big| \dn{n},G) = f_{\cal PD}\big( \dn{\pi}^G; \, \widetilde{\dn{\alpha}}^G \big)$
with
$$
\widetilde{\dn{\alpha}}^G=\Big( \widetilde{\dn{\alpha}}_d, \, d \in {\cal D}(G) \Big) \mbox{~~and~~}
\widetilde{\dn{\alpha}}_d = \Big( \widetilde{\alpha}_d(i_d) = \alpha_d(i_d)+n_d(i_d), ~i_d \in {\cal I}_d\Big).
$$
Finally, for the gamma structure
\mbox{$f( \dn{\pi}^G|\dn{n},G )=  f_{\cal CPD} \big( \dn{\pi}_{c|\overline{c}} \, ; \, \widetilde{\dn{\alpha}}  \big)
\times f_{\cal PD } \big( \dn{\pi}_d ; \,\widetilde{\dn{\alpha}}_d, \, d\in{\cal D}(G) \big)$}
i.e. a distribution with density equivalent to the corresponding prior (\ref{prior_gamma}) with parameters
$$
\widetilde{\dn{\alpha}}^G
=\Big( \widetilde{\dn{\alpha}}, \,
       \widetilde{\dn{\alpha}}_d, \, d \in {\cal D}(G)\Big) .
$$

>From the properties of the Dirichlet distribution, it derives that
each element of $ \dn{\pi}^G$ follows a Beta distribution with the
appropriate parameters.

 For model choice we need  to estimate the  posterior model
probabilities $f(G|n)\propto f(\dn{n}|G)f(G)$, with $f(\dn{n}|G)$ marginal likelihood of the model
and $f(G)$ prior distribution on $G$. Here we restrict to the simple
case where $f(G)$ is uniform, hence the
posterior will depend only on the marginal likelihood
$f(\dn{n}|G)$ of the model under consideration. The marginal
likelihood can be calculated analytically since the above prior
set-up is conjugate.

For the saturated model the marginal likelihood is given by
$$
f(\dn{n}|G) =  K( \dn{n} )\times \frac{ DK\big( \dn{\alpha}  \big) }{DK\big( \widetilde{\dn{\alpha}} \big)}
$$
where $K( \dn{n} ) $ and $DK( \dn{\alpha} )$ are given by
$$
K( \dn{n} ) =  \frac{\Gamma(N+1)}{\prod \limits_{i \in {\cal I}}\ \Gamma\Big(n(i)+1\Big)}
\mbox{~~and~~}
DK\big( \dn{\alpha} \big) =\frac{\Gamma\left(\sum \limits_{i \in {\cal I}}\ \alpha(i)\right)}
     {\prod \limits_{i \in {\cal I}}\Gamma\Big(\alpha(i)\Big)}~.
$$
respectively.

For the independence and the edge models the marginal likelihood is given by
\begin{eqnarray}
 f(\dn{n}|G)= K( \dn{n} ) \prod_{ d \in {\cal D}(G)}
 \frac{ DK\big( \dn{\alpha}_d  \big) }
      { DK\big( \widetilde{\dn{\alpha}}_d  \big)}
\end{eqnarray}
where
$$
DK\big( \dn{\alpha}_d  \big) = \frac{\Gamma\left(\sum \limits_{i_d \in {\cal I}_d}\ \alpha_d(i_d)\right)}
                                    {\prod \limits_{i_d \in {\cal I}_d}\Gamma\Big(\alpha_d(i_d)\Big)}~.
$$
Finally, for the gamma structure the marginal likelihood $f(\dn{n}|G)$ is given by
\begin{eqnarray}
f(\dn{n}|G) &=&
K( \dn{n} )
 \prod_{i_{\overline{c} }\; \in {\cal I}_{\overline{c} }}
 \frac{ DK\big( \dn{\alpha}( \cdot ,i_{\overline{c}})  \big) }
      { DK\big( \widetilde{\dn{\alpha}}(\cdot,i_{\overline{c}}) \big) }
  \prod_{ d \in {\cal D}(G)}
  \frac{ DK\big( \dn{\alpha}_d \big) }
      { DK\big( \widetilde{\dn{\alpha} }_d \big)}~.
\end{eqnarray}
where
$$
DK\Big( \dn{\alpha}( \cdot ,i_{\overline{c}}) \Big) = \frac{\Gamma\left(\sum \limits_{i_c \in {\cal I}_c}\ \alpha(i_c, i_{\overline{c}})\right)}
            {\prod \limits_{i_c \in {\cal I}_c}\Gamma\Big(\alpha(i_c, i_{\overline{c}})\Big)}~.
$$

\medskip

The posterior distribution of the marginal log-linear parameters
$\dn{\lambda}^G$ can be estimated in a straightforward manner
using  Monte Carlo samples from the posterior distribution of
 $\dn{\pi}^G$.
Specifically, a  sample from the posterior distribution of
$\dn{\lambda}^G$ can be generated by the following steps.
\begin{enumerate}
\item[i)] Generate a random sample $\dn{\pi}^{G,\,(t)} (t=1,\dots,T)$ from the posterior distribution of   $\dn{\pi}^G$.
\item[ii)] At each iteration $t$, calculate the the full table of probabilities $\dn{\pi}^{(t)}$ from $\dn{\pi}^{G,\,(t)}$.
\item[iii)] The vector of marginal log linear parameters,  $\dn{\lambda}^{G,\,(t)}$, can be easily obtained from
          $\dn{\pi}^{(t)}$ via equation (\ref{lambda_of_marginal_model}) which becomes
          $$
          \dn{\lambda}^{G,\,(t)} = \dn{C}^G \log \Big( \dn{M}^G \vet \big(\dn{\pi}^{(t)}\big) \Big)
          $$
          where $\dn{C}^G$ and  $\dn{M}^G$ are the contrast and marginalization matrices under graph $G$.
          Note that some elements of $\dn{\lambda}^G$ will automatically be constrained to zero for all
          generated values due to the graphical structure of the model $G$ and the way we calculate
          log-linear parameters using the previous equation.
\end{enumerate}
Finally, we can use the generated values $\big(
\dn{\lambda}^{G,\,(t)}; t=1,2,\dots,T \big)$ to estimate summaries
of the posterior distribution $f(\dn{\lambda}^{G}|G)$ or obtain
plots fully describing this distribution.

\section{Illustrative examples}

The methodology described in the previous sections is now
illustrated on two real data sets, a $2 \times 2 \times 2 $ and a
$3 \times 2 \times4 $  tables. In both example we compare the
results obtained with our yardstick prior, the UIP-Perks' prior ($\alpha(i)
= 1/|I|$), with those obtained using Jeffrey's ($\alpha(i) = 1/2$),
Unit Expected Cell ($\alpha(i) = 1$), and Empirical Bayes
($\alpha(i) = p(i)$)
 priors.

\subsection{A $2\times 2 \times 2$ Table: Antitoxin Medication Data}

We consider a data set presented by Healy (1988) regarding a study
on the relationship between patient condition (more or less
severe), assumption of antitoxin (yes or not) and survival status
(survived or not); see  Table \ref{log-reg-exam}. In Table \ref{Antitoxin} we compare posterior model
probabilities
under the
 four different prior set-ups.

\begin{table}[h]
\begin {center}
\caption [log-dat1] {Antitoxin data}\vspace{0.2cm} \label
{log-reg-exam}
\begin {tabular} {l c c c} \hline
 &  & \multicolumn {2}{c}{Survival (S)}\\ \hline
Condition (C) & Antitoxin (A) & No & Yes\\
\hline
More Severe & Yes & 15 & 6 \\
 & No & 22 & 4 \\
\hline
Less Severe & Yes & 5 & 15 \\
 & No & 7 & 5 \\
\hline
\end {tabular}
\end {center}
\end{table}

Under all prior assumptions the maximum a posteriori  model (MAP)
is  SC+A (we omit the conventional crossing (*) operator
between variables for simplicity), assuming  the
marginal independence of  Antitoxin from the remaining two
variables.

Under Empirical Bayes and UIP-Perks' priors the posterior
distribution is concentrate on the MAP model (it takes into
account $93.4 \%$ and $91.7 \%$ respectively of the posterior
model probabilities). The posterior distributions under the
Jeffreys' and the unit expected prior set-ups are more disperse,
supporting the three models (SC+A, AS+SC and ASC) with posterior
weights higher than $10\%$ and accounting around the  $94\%$ of
the posterior model probabilities. Model $AS+SC$ is also the model
with the se\-cond highest posterior probability under UIP-Perks'
prior but its weight is considerably lower than the corresponding
probability of the MAP model.

\begin{table}[h]
\caption{ Posterior model probabilities ($\%$) for the Antitoxin
data.
}\vspace{0.2cm} \label{Antitoxin} {\scriptsize \centering
\begin{tabular}{l@{~~}l@{~~~}|c@{~~~}c@{~~~}c@{~~~}c@{~~~}c@{~~~}c@{~~~}c@{~~~}c}
\hline
                        &                   & \multicolumn{8}{c}{ Model } \\
                        &                   & A+S+C & AS+C & AC+S & SC+A & AS+AC &
AS+SC & AC+SC &
ASC\\
Prior Distribution      &                   &  (1) & (2) & (3) & (4)  & (5) & (6) &
(7) & (8) \\
\hline
Jeffreys'          & $\alpha(i) = 1/2$    & 0.3 & 1.5 & 0.2 & 59.7 & 0.1 & 21.7 &
3.0 & 13.4 \\
Unit Expected Cell & $\alpha(i) = 1$      & 0.2 & 1.1 & 0.2 & 37.2 & 0.1 & 30.2 &
4.7 & 26.2\\
Empirical Bayes    & $\alpha(i) = p(i)$   & 1.6 & 2.4 & 0.3 & 93.4 & 0.0 &~~1.7 &
0.2 &~~0.4\\
UIP-Perks'             & $\alpha(i) = 1/|I|$  & 1.2 & 2.1 & 0.3 & 91.7 & 0.0 &~~3.5
& 0.4 &~~0.8 \\
\hline
\end{tabular}
\normalsize } \label{models}
\end{table}

\begin{figure}[hbtp]
      \caption{ Antitoxin data: Boxplots summarizing 2.5\%, 97.5\% posterior
percentiles and
quantiles of the joint probabilities $\pi_{ABC}(i,j,k)$
          for the MAP model (SC+A)  for all prior
          set-ups (J=Jeffreys',
          U=Unit Expected Cell, E=Empirical Bayes, P=Perks') .
          \label{ex1boxplot2}}
      \vspace{-1cm}
      \begin{center}
      \includegraphics[width=12cm]{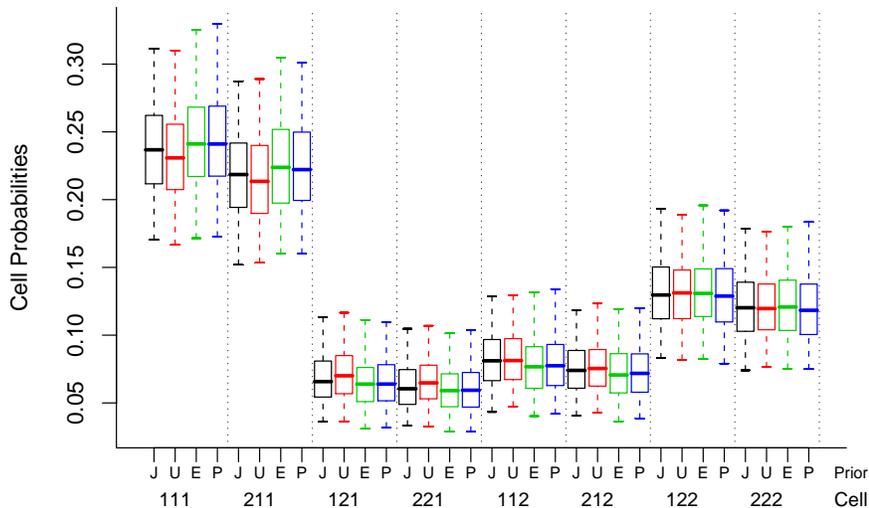}
\vspace{-1cm}
      \end{center}
   \end{figure}

Figure  \ref{ex1boxplot2} presents boxplots summarizing 2.5\%,
97.5\% posterior percentiles and quantiles of the joint
probabilities  for the MAP model (SC+A) for the four prior
set-ups. Since direct calculation from the posterior distribution
is not feasible, we estimated the posterior summaries via  Monte
Carlo simulation (1000 values). From this figure, we observe minor
differences between the posterior distributions obtained under the
UIP-Perks' and the empirical Bayes prior. More differences are
observed between Perks' UIP and the posterior distributions under
the two other prior set-ups. Differences are higher for the first
two cell probabilities, i.e. for $\pi(1,1,1)$ and $\pi(2,1,1)$.

\begin{figure}[hbtp]
      \caption{Antitoxin data: Boxplots summarizing 2.5\%, 97.5\% posterior
percentiles and
quantiles of the joint probabilities $\pi_{ABC}(i,j,k)$
          for models SC+A, AS+SC and ASC (4, 6 and 8 respectively)  under the
UIP-Perks' prior
          set-up.
           \label{ex1boxplot}}
      \vspace{-2cm}
      \begin{center}
      \includegraphics[width=12cm]{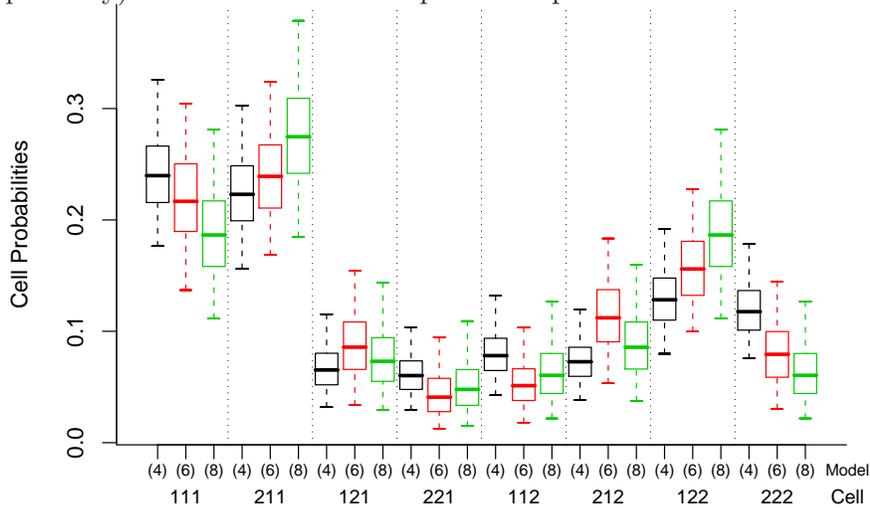}
      \end{center}
   \end{figure}

Similarly in Figure \ref{ex1boxplot} we present boxplots providing
posterior summaries for models $SC+A$, $AS+SC$ and $ASC$ under the
UIP-Perks' prior set-up. The first two models are the ones with
highest posterior probabilities and all of their summaries have
been calculated using Monte Carlo simulation (1000 values). The
saturated was used mainly as reference model since it is the only
one for which the posterior distributions are available
analytically. From the figure we observe that the posterior
distributions on the joint probabilities $\dn{\pi}$ of the full
table are quite different highly depending on the assumed model
structure.

Finally, posterior summaries for the probability parameters $\dn{\pi}^G$
and the marginal log-linear parameters $\dn{\lambda}^G$
for models $SC+A$, $AS+SC$ and $ASC$ (as described above) under the UIP-Perks'  prior
are provided in Tables \ref{Antitoxin_post_pis} and \ref{Antitoxin_post_lambdas}
respectively.
All summaries of each element of $\dn{\pi}^G$ are obtained analytically based on the
Beta
distribution
induced by the corresponding Dirichlet posterior distributions of $\dn{\pi}^G$.
Posterior summaries of $\dn{\lambda}^G$ are estimated using the Monte Carlo strategy
(1000 values)
discussed
in section \ref{post_distributions}.
As commented in this section, some elements of $\dn{\lambda}^G$ for
graphs $SC+A$ and $AS+SC$ are constrained to zero due the way we have constructed
our model.
Hence for $SC+A$, the maximal interaction terms for the disconnected sets $AS$, $AC$
and $ASC$,
i.e. parameters $\lambda_{AS}(2,2)$, $\lambda_{AC}(2,2)$ and $\lambda_{ASC}(2,2,2)$,
are constrained to be zero for all generated observations.
Similar is the picture for model $AS+SC$, but now only marginals $AC$ and $ABC$
correspond to
disconnected sets
implying that $\lambda_{AC}(2,2)=\lambda_{ASC}(2,2,2)=0$.

\begin{table}[p]
\begin{center}
\caption{ Posterior summaries of model parameters for models $SC+A$, $AS+SC$ and $ASC$
in Antitoxin data under the UIP-Perks' prior set-up;
$\widetilde{a}$ and $\widetilde{b}$ are the parameters of the resulted Beta marginal
posterior
distribution. }
\vspace{0.2cm} \label{Antitoxin_post_pis} {\small
\centering
\begin{tabular}{l|cccccc}
\hline
\multicolumn{2}{l}{Model 4: $SC+A$ }&    &   &   &  \multicolumn{2}{c}{Beta Posterior}\\
 &    &    &   &   &  \multicolumn{2}{c}{Parameters}\\
Parameter &  Mean  &  St.dev.  &  $Q_{0.025}$  &  $Q_{0.975}$  &  $\widetilde{a}$  & $\widetilde{b}$ \\
\hline
$\pi_{SC}(1,1)$  &  0.47  &  0.055  &  0.36  &  0.57  &  37.25  &  42.75 \\
$\pi_{SC}(2,1)$  &  0.13  &  0.037  &  0.06  &  0.21  &  10.25  &  69.75 \\
$\pi_{SC}(1,2)$  &  0.15  &  0.040  &  0.08  &  0.24  &  12.25  &  67.75 \\
$\pi_A(1)$       &  0.52  &  0.056  &  0.41  &  0.63  &  41.50  &  38.50 \\
\hline \hline
\multicolumn{2}{l}{Model 6: $AS+SC$} \\
Parameter &  Mean  &  St.dev.  &  $Q_{0.025}$  &  $Q_{0.975}$  &  $\widetilde{a}$  &
$\widetilde{b}$ \\
\hline
$\pi_{S|AC}(1|1,1)$  &  0.71  &  0.096  &  0.51  &  0.88  &  15.12  & ~~6.12\\
$\pi_{S|AC}(1|2,1)$  &  0.84  &  0.070  &  0.68  &  0.95  &  22.12  & ~~4.12\\
$\pi_{S|AC}(1|1,2)$  &  0.25  &  0.094  &  0.09  &  0.46  & ~~5.12  &  15.12\\
$\pi_{S|AC}(1|2,2)$  &  0.58  &  0.136  &  0.31  &  0.83  & ~~7.12  & ~~5.12\\
$\pi_A(1)$            &  0.52  &  0.056  &  0.41  &  0.63  &  41.50  &  38.50\\
$\pi_C(1)$            &  0.59  &  0.055  &  0.48  &  0.70  &  47.50  &  32.50\\
\hline \hline
\multicolumn{4}{l}{Model 8: $ASC$ (Saturated)}\\
Parameter &  Mean  &  St.dev.  &  $Q_{0.025}$  &  $Q_{0.975}$  &  $\widetilde{a}$  &
$\widetilde{b}$ \\
\hline
$\pi(1,1,1)$  &  0.19  &  0.044  &  0.11  &  0.28  &  15.12  &  64.88\\
$\pi(2,1,1)$  &  0.28  &  0.050  &  0.18  &  0.38  &  22.12  &  57.88\\
$\pi(1,2,1)$  &  0.08  &  0.030  &  0.03  &  0.14  & ~~6.12  &  73.88\\
$\pi(2,2,1)$  &  0.05  &  0.025  &  0.01  &  0.11  & ~~4.12  &  75.88\\
$\pi(1,1,2)$  &  0.06  &  0.027  &  0.02  &  0.13  & ~~5.12  &  74.88\\
$\pi(2,1,2)$  &  0.09  &  0.032  &  0.04  &  0.16  & ~~7.12  &  72.88\\
$\pi(1,2,2)$  &  0.19  &  0.044  &  0.11  &  0.28  &  15.12  &  64.88\\
$\pi(2,2,2)$  &  0.06  &  0.027  &  0.02  &  0.13  & ~~5.12  &  74.88\\
\hline
\end{tabular}
\normalsize
}
\end{center}
\end{table}

\small \tabspace \normalsize

\begin{table}[p]
\begin{center}
\caption{Antitoxin data: Posterior summaries for lambda for models
SC+A, AS+SC and ASC  under the UIP-Perks' prior set-up
}\vspace{0.2cm}  \label{Antitoxin_post_lambdas}{\small \centering
\begin{tabular}{l|ccccc}
\hline
\multicolumn{2}{l}{Model 4:  $SC+A$ }\\
Parameter  & Marginal table& Mean  &  St.dev.  &  $Q_{0.025}$  &  $Q_{0.975}$   \\
\hline
$\lambda_{\emptyset}$   & $M_{AS}$  &   -1.429  & 0.032 & -1.513& -1.388\\
$\lambda_{A}(2)$ & $M_{AS}$  &      -0.040 &  0.113& -0.258 & 0.181\\
$\lambda_S(2)$ &$M_{AS}$&      -0.245 &  0.118 & -0.480 & -0.021\\
$\lambda_{AS}(2,2)$  &$M_{AS}$&  0.000 &   0.000 & 0.000  & 0.000\\
$\lambda_{C}(2)$     &$M_{AC}$&  -0.194 &  0.116 & -0.426 &  0.027\\
$\lambda_{AC}(2,2)$   & $M_{AC}$&  0.000  & 0.000 & 0.000 & 0.000\\
$\lambda_{SC}(2,2)$  &$M_{ASC}$&  0.460 &  0.134 & 0.199 & 0.735\\
$\lambda_{ASC}(2,2,2)$ &$M_{ASC}$& 0.000 &  0.000 & 0.000 & 0.000\\
\hline \hline
\multicolumn{2}{l}{Model 6: $AS+SC$} \\
Parameter  & Marginal table& Mean  &  St.dev.  &  $Q_{0.025}$  &  $Q_{0.975}$   \\
$\lambda_{\emptyset}$    & $M_{AC}$  &        -1.418 &  0.025 & -1.483 & -1.388\\
$\lambda_A(2)$   & $M_{AC}$&     -0.042  & 0.114 & -0.261 &  0.173\\
$\lambda_C(2)$      & $M_{AC}$& -0.195  & 0.110  & -0.414 & 0.020\\
$\lambda_{AC}(2,2)$   & $M_{AC}$&  0.000 &  0.000 & 0.000 & 0.000\\
$\lambda_{S}(2)$ &$M_{ASC}$ &     -0.238 &  0.137 & -0.493 &  0.044\\
$\lambda_{AS}(2,2)$  &$M_{ASC}$ &-0.291  & 0.137 & -0.554 & -0.019\\
$\lambda_{SC}(2,2)$  & $M_{ASC}$ & 0.437 &  0.137 & 0.178 & 0.712\\
$\lambda_{ASC}(2,2,2)$ &$M_{ASC}$ &-0.086  & 0.143 & -0.370 & 0.207\\
\hline \hline
\multicolumn{4}{l}{Model 8: $ASC$ (Saturated)}\\
Parameter  & Marginal table& Mean  &  St.dev.  &  $Q_{0.025}$  &  $Q_{0.975}$   \\
$\lambda_{\emptyset}$      &$M_{ASC}$ & -2.325  & 0.079 & -2.504 & -2.191\\
$\lambda_A(2)$   &$M_{ASC}$ &  -0.106 &  0.134 &-0.379 & 0.152\\
$\lambda_S(2)$    &$M_{ASC}$ & -0.246 &  0.131 &-0.510 & 0.004\\
$\lambda_{AS}(2,2)$ & $M_{ASC}$ &  -0.292  & 0.139 &-0.576 &-0.033\\
$\lambda_{C}(2)$    &$M_{ASC}$ & -0.136 &  0.143 &-0.402 & 0.151\\
$\lambda_{AC}(2,2)$  &$M_{ASC}$ & -0.084 &  0.139 & -0.355 & 0.202\\
$\lambda_{SC}(2,2)$  &$M_{ASC}$ &  0.450  & 0.135 & 0.207 & 0.705\\
$\lambda_{ASC}(2,2,2)$ &$M_{ASC}$ & -0.074  & 0.143 &-0.368 & 0.209\\
\hline \hline
\end{tabular}
\normalsize }
\end{center}
\end{table}

\small \tabspace \normalsize

\subsection{A $3 \times 2 \times 4$ table: Alcohol Data}

 We now examine  a  well known  data set presented by Knuiman and Speed
(1988) regarding  a small study held in Western Australia on the
relationship between Alcohol intake (A), Obesity (O) and High
blood pressure (H); see Table \ref{tabc}.

\begin{table}[hbtp]
\begin{center}
\caption{\label{tabc} Alcohol Data} \vspace{0.2cm}
\begin{tabular}{llcccccc}
\hline &  & \multicolumn{5}{c}{Alcohol intake} \\
\hline &  & \multicolumn{5}{c}{ (drinks/days)} \\ \cline{3-7}
Obesity & High BP  &  & ~0 & 1-2 & 3-5 & 6+ \\  \hline
Low     & Yes      &  & ~5 & ~9  & ~8  & 10 \\
        & No       &  & 40 & 36  & 33  & 24 \\
Average & Yes      &  & ~6 & ~9  & 11  & 14 \\
        & No       &  & 33 & 23  & 35  & 30 \\
High    & Yes      &  & ~9 & 12  & 19  & 19 \\
        & No       &  & 24 & 25  & 28  & 29 \\ \hline
\end{tabular}
\end{center}
\end{table}

In Table \ref{alcohol_log.like} we report  posterior model probabilities  and corresponding Log-marginal likelihoods for each models.
Under all prior set-ups the posterior model probability is concentrated on models H+A+O, HA+O and HO+A.
Empirical Bayes and UIP-Perks' support the independence model (with posterior model probability
of $0.878$ and $0.807$ respectively) whereas Jeffreys' and Unit Expected support a more complex structure,
HO+A (with posterior model probability
of $0.837$ and $0.859$ respectively).

\begin{table}[hp!]
\begin{center}
\caption{Alcohol data: Posterior model probabilities and the corresponding Log marginal likelihoods;
 empty cell in posterior probabilities means that it is lower than $0.0001$
}\vspace{0.2cm} \label{alcohol_log.like}
{\scriptsize
\begin{tabular}{l@{~}|c@{~~~}c@{~~~}c@{~~~}c@{~~~}c@{~~~}c@{~~~}c@{~~~}c}
\hline\hline \multicolumn{1}{l}{~} \\
\multicolumn{9}{l}{ Posterior model probabilities ($\%$)} \\
\cline{1-3}
                        & \multicolumn{8}{c}{ Model } \\
                        & H+A+O & HA+O & HO+A & AO+H & HA+HO & HA+AO & HO+AO & HAO \\
Prior Distribution      &  (1) & (2) & (3) & (4)  & (5) & (6) & (7) & (8) \\
\hline
Jeffreys'            &  11.56  &  4.76  &  83.68   \\
Unit Expected Cell   & ~~6.91  &  7.21  &  85.88   \\
Empirical Bayes      &  87.81  &  0.07  &  12.12   \\
Perks'               &  80.67  &  0.15  &  19.18   \\
\hline\hline \multicolumn{1}{l}{~} \\
\multicolumn{9}{l}{ Log-marginal likelihood for each model } \\
\cline{1-3}
                        & \multicolumn{8}{c}{ Model } \\
                        & H+A+O    & HA+O & HO+A & AO+H & HA+HO & HA+AO & HO+AO & HAO \\
Prior Distribution      &  (1) & (2) & (3) & (4)  & (5) & (6) & (7) & (8) \\
                        \hline
Jeffreys           ($\alpha(i)=1/2$)    &  -79.22  &  -80.11  &  -77.24  & ~~-87.73  & ~~-90.44  &  -100.93  & ~~-98.06  & ~~-98.95   \\
UEC                ($\alpha(i)=1$)      &  -78.51  &  -78.47  &  -75.99  & ~~-84.70  & ~~-85.27  & ~~-93.99  & ~~-91.51  & ~~-91.46   \\
Emprirical Bayes   ($\alpha(i)=p(i)$)  &  -86.96  &  -94.10  &  -88.94  &  -107.26  &  -124.75  &  -143.06  &  -137.91  &  -145.04   \\
Perks              ($a(i)=1/|I|$)  &  -86.90  &  -93.19  &  -88.33  &  -107.10  &  -121.13  &  -139.89  &  -135.03  &  -141.33   \\
\hline\hline
\end{tabular}
\normalsize
}
\end{center}
\end{table}

To save space we do not report here posterior summaries for model
parameters, they can be found in a separate appendix on  the
web page:
\begin{center}
\verb|http://stat-athens.aueb.gr/~jbn/papers/paper21.htm|.
\end{center}

\section{Discussion and Final Comments}

In this paper we have dealt with the Bayesian analysis of graphical models of marginal
association  for three way contingency tables. We have
worked using the probability parameters of marginal tables
required to fully specify each model. The proposed parametrization
and the corresponding decomposition of the likelihood simplifies
the problem and automatically imposes the  marginal independences
represented by the considered graph. By this way, the posterior
model probabilities and the posterior distributions for the used
parameters can be calculated analytically. Moreover, the posterior
distributions of the marginal log-linear parameters
$\dn{\lambda}^G$ and the probabilities $\dn{\pi}$ of the full
table can be easily obtained using simple Monte Carlo schemes.
This approach avoids the problem of the inverse calculation of
$\dn{\pi}$ when the marginal association log-linear parameters
$\dn{\lambda}$ are available which can be only achieved via an
iterative procedure; see Rudas and Bergsma (2004) and Lupparelli (2006) for more details.

An obvious extension of this work is to implement the same
approach in tables of higher dimension starting from four way tables.
Although most of the models in a four way contingency table can be factorized and analyzed in a similar manner,
two type of graphs (the 4-chain and the cordless
four-cycle graphs) cannot be decomposed in the above way.
These models are not Markov equivalent to any  directed acyclic graph (DAG).
In fact each bidirected graph (which corresponds to a marginal
association model) is equivalent to a DAG, i.e. a conditional
association model, with the same set of variables if and only if
it does not contain any 4-chain, see Pearl and Wermuth (1994). We
believe that also in higher dimensional problems our approach can be applied to bidirected
graphs that admit a DAG representation.
For the graph that do not factorize, more sophisticated techniques must be adopted in order to
obtain the posterior distribution of interest and the
corresponding marginal likelihood needed for the model comparison
(work in progress by the authors).

%

Another interesting subject is how to obtain the posterior
distributions in the case that someone prefers to work directly
with  marginal log-linear parameters $\dn{\lambda}^G$ defined by
(\ref{lambda_of_marginal_model}). Using our approach, we impose a
prior distribution on the probability parameters $\dn{\pi}^G$. The
prior of $\dn{\lambda}^G$ cannot be calculated analytically since
we cannot have the inverse expression of
(\ref{lambda_of_marginal_model}) in closed form. Nevertheless,  we
can obtain  a sample from the imposed prior on
$\dn{\lambda}^G$ using a simple Monte Carlo scheme. More
specifically, we can generate random values of $\dn{\pi}^G$ from
the Dirichlet based prior set-ups described in this paper.
We calculate the joint probability vector $\dn{\pi}$ according to the
factorization of the graph under consideration and finally use
(\ref{lambda_of_marginal_model}) to obtain a sample from the
imposed prior $f(\dn{\lambda}^G|G)$. This will give us an idea of
the prior imposed on the log-linear parameters.

If prior information is  expressed directly in terms of the   log-linear parameters,
see e.g Knuiman and
Speed (1988) and Dellaportas and Forster (1999),    the prior
and the corresponding posterior distribution of $\dn{\pi}^G$ can be
obtained  using two alternative strategies.

One possibility is to approximate the distribution imposed on the elements
of $\dn{\pi}^G$ via Dirichlet distributions with the parameters
obtained in the following way.
Firstly we generate random values from the prior imposed on the standard log-linear
parameters for models of conditional association.
For each set of generated  values, we calculate
the corresponding probabilities $\dn{\pi}$ for the full table.
Finally we obtain a sample for $\dn{\pi}^G$ via marginalization from each set of generated probabilities $\dn{\pi}$.
For every element of $\dn{\pi}^G$,
we  use the corresponding generated values to approximate the imposed prior by a Dirichlet distribution with the parameters
estimated using the moment-matching approach.
Note that this approach can only provide us a rough picture of the correct
posterior distribution since the priors are only matched in terms
of the mean and the variance while their shape can be totally
different due to the properties of the Dirichlet distribution.

Similar will be the approach if the prior distributions $f(\dn{\lambda}^G|G)$
for the marginal log-linear parameters $\dn{\lambda}^G$ are available.
The only problem here, in comparison to the
simpler approach described in the previous paragraph, is the
calculation of $\dn{\pi}$ from each $\dn{\lambda}^G$. In order to
achieve that we need to use iterative procedures; see  Rudas and Bergsma (2004) and Lupparelli (2006).

A second approach  is to directly calculate the prior distribution imposed
on the probability parameters $\dn{\pi}^G$ starting from the prior
$f(\dn{\lambda}^G|G)$ using equation
(\ref{lambda_of_marginal_model}). Note that the probabilities
$\dn{\pi}$ of the full table involved in
(\ref{lambda_of_marginal_model}) are simply a function of
$\dn{\pi}^G$ depending on the structure $G$. Hence, the prior on
$\dn{\pi}$
will be given by
$$
f(\dn{\pi}^G|G) = f(\dn{\lambda}^G|G) \left|\frac{\partial
\dn{\lambda}^G}{\partial \dn{p}^G} \right|
$$
where $\dn{p}^G$ is $\mbox{vec}(\dn{\pi}^G)$  after removing the
last element of each set of probability parameters
and $\mbox{vec}(\dn{\pi}^G)$ refers to $\dn{\pi}^G$ arranged in a
vector form. The elements of the Jacobian are given by
$$
\frac{\partial \lambda^G_k}{\partial  p_l^G } = \sum_{i=1}^{
col(\mbox{\footnotesize \bf C})} \left\{ C_{ki} \left(
\sum_{j=1}^{|{\cal I}|} M_{ij} \mbox{vec}({\dn{\pi}})_j
\right)^{-1}
              \sum_{j=1}^{|{\cal I}|} M_{ij} \frac{\partial \mbox{vec}({\dn{\pi}})_j}
                                                  {\partial p_l^G } \right\}
$$
where $col(\mbox{\footnotesize \bf C})$ is the number of columns
of $\dn{C}$ matrix. For the saturated model the above equation
simplifies to
$$
\frac{\partial \lambda^G_k}{\partial p_l }  = \sum_{i=1}^{
col(\mbox{\footnotesize \bf C}) } \frac{ C_{ki}  ( M_{il} -
M_{i|{\cal I}|} ) } { \sum_{j=1}^{|{\cal I}|} M_{ij}
\mbox{vec}({\dn{\pi}})_j }
$$
since $\mbox{vec}(\dn{\pi})_j = p_j^G$ for $j < |{\cal I}|$ and
$\mbox{vec}(\dn{\pi})_{|{\cal I}|} = 1 - \sum_{j=1}^{|{\cal I}|-1}
p_j^G$. After calculating the corresponding prior distribution
$f(\dn{\pi}^G|G)$, we can work directly on $\dn{\pi}^G$ using an
MCMC algorithm to generate values from the resulted posterior. A
sample of $\dn{\lambda}^G$ can be again obtained in a direct way
using (\ref{lambda_of_marginal_model}). When  no strong prior
information is available, an independence Metropolis algorithm can
be applied using as a proposal the Dirichlet distributions
resulted from the likelihood part. Otherwise more sophisticated
techniques might be needed. The authors are also exploring the
possibility to extend the current work in this direction.

\section*{Acknowledgment}
We thank Monia Lupparelli for fruitful discussions and Giovanni
Marchetti for providing us the {\tt R} function  {\tt inv.mlogit}.
This work was partially supported by MIUR, ROME, under project
PRIN 2005132307 and University of Pavia.

\section*{Appendix}


\begin{enumerate}
\item{\textbf{ Construction of Matrix $\mathbf M$}}

Let ${\cal M}=\{ M_1, M_2, \dots, M_{|{\cal M}|} \}$ be the set of
considered marginals. Let  ${\mathbf B}$ be a  binary matrix of
dimension $|\cal M| \times |{\cal V}|$ with elements $B_{iv}$
indicating whether a variable $v$ belongs to a specific marginal
${M}_i$. The rows of ${\mathbf B}$ correspond to the
 marginals in ${\cal M}$ whereas the columns to the variables. The variables follow a reverse
 ordering, that is  column 1  corresponds to variable $X_{|\cal V|}$, column 2 to variable
  $X_{|{\cal V}|-1}$ and so on.
Matrix $\mathbf   B$ has elements
$$
B_{iv}=\left\{
\begin{array}{rl}
   1 &  \mbox{ if   $v \in { M}_i$ } \\
   0 &  \mbox{ otherwise.}
\end{array}
\right.,
$$
for every $v \in \cal V$.

The marginalization matrix $\mathbf M$ can be constructed using
the following rules.
\begin{enumerate}
\item For each marginal ${M}_i$, the probability vector of the
corresponding marginal table is given by $\mathbf M_i \mathbf
\pi$;
            where $\mathbf M_i$ is calculated as a Kronecker product of matrices ${\mathbf
            A}_{iv}$
        $$
        \mathbf   M_i = \bigotimes _{v \in {\cal V}} {\mathbf   A}_{iv}
        $$
        with
$$
\mathbf   {\mathbf   A}_{iv}=\left\{
\begin{array}{rl}
   {\mathbf   I}_{\ell_v} & \mbox{ if  }\; \;  B_{iv} = 1\\
   {\mathbf   1}_{\ell_v}^T &\mbox{ if  } \; \;B_{iv} = 0
\end{array}
\right.
$$
        where
        $\ell_v$ is the number of levels for $v$ variable,
        ${\mathbf   I}_{\ell_v}$ is the identity matrix of dimension $\ell_v\times \ell_v$ and
        ${\mathbf   1}_{\ell_v}$ is a vector of dimension $\ell_v \times 1$ with all elements equal to one.
\item Matrix $\mathbf   M$ is constructed by stacking all the
$\mathbf M_i$ matrices
                $$
        \mathbf M = \left(  \begin{array}{l} \mathbf M_1 \\ \vdots \\ \mathbf M_i  \\ \vdots \\ \mathbf M_{|\mathbb{{ M}}|} \end{array}  \right)
        $$
\end{enumerate}

\

\item \textbf{Construction of  Matrix $\mathbf C$}

Firstly we need to construct the design matrix $X$ for the
saturated model corresponding to sum to zero constraints. It has
has dimension $\left( \prod \limits _{v} \ \ell_v \right)\times
\left( \prod \limits _{v} \ \ell_v \right) $ and  can be obtained
as
$$
\mathbf X =\underset{v \in {\cal V}^R} \bigotimes {\mathbf
J}_{\ell_v}
$$
with
$${\mathbf   J}_{\ell_v}(r,c)=\left\{
\begin{array}{rl}
   ~1 &  \mbox{ if  $c=1$ or $r=c$}  \\
   -1 &  \mbox{ if  $r=1$ and  $c>1$} \\
   ~0 &     \mbox{ otherwise.}
\end{array}
\right.$$ In matrix notation
$${\mathbf   J}_{\ell_v}=\left(
\begin{array}{ll}
        1             & - \mathbf{1}^T_{(\ell_v-1)} \\
\mathbf{1}_{(\ell_v-1)} & \mathbf{I}_{(\ell_v-1)\times(\ell_v-1)}
\end{array}\right)
$$

where $\mathbf 1_{(\ell_v-1)}$ is $({\ell_v-1}) \times 1$ vector
of ones while $\mathbf I_{({\ell_v-1})\times ({\ell_v-1})}$ is an
identity matrix of dimension $({\ell_v-1})\times ({\ell_v-1})$.

%
%
%
%
%

The contrast matrix  $\mathbf C$ can be constructed by  using the
following rules.

\begin{enumerate}
\item For each margin ${ M}_i$  construct the design matrix
$X_i$ corresponding to the saturated model (using  sum to zero
constraints)  and invert it to get the contrast matrix for the
saturated model $\mathbf C_i=X^{-1}$ . Let $\mathbf C_i^*$ be a
submatrix of $\mathbf C_i$ obtained by deleting rows not
corresponding to elements of $\dn{\cal E}_{M_i}$ (the  effects
that we wish to estimate from margin ${ M}_i $) .

\item The contrast matrix   $\mathbf C$ is obtained by direct sum
of the $C_i^*$ matrices as follow
$$
\mathbf C = \bigoplus_{i:~ {\cal M}_i \in {\mathbb M}}  \mathbf
C_i^*
$$
that is it is a block diagonal matrix with $( \mathbf C_1^* ;
{ M}_i \in {\cal M} )$ as the blocks. For example $C_i^*
\bigotimes C_2=\bigoplus_{i=1}^{2}C^*_i$ is the block diagonal
matrix with $C_1$ and $C_2$ as blocks.
\end{enumerate}


\end{enumerate}

\end{document}